\documentclass[10pt,conference]{IEEEtran}
\IEEEoverridecommandlockouts
\usepackage{cite}
\usepackage{amsmath,amssymb,amsfonts}
\usepackage{algorithmic}
\usepackage{graphicx}
\usepackage{textcomp}
\usepackage{xcolor}
\usepackage{balance}
\usepackage{tcolorbox}
\usepackage{blindtext}
\usepackage{enumitem}
\usepackage{xspace}
\usepackage{url}
\usepackage{booktabs}
\usepackage{xcolor}

\def\BibTeX{{\rm B\kern-.05em{\sc i\kern-.025em b}\kern-.08em
    T\kern-.1667em\lower.7ex\hbox{E}\kern-.125emX}}

\newcommand{\ie}{\textit{i.e.,}\xspace}
\newcommand{\eg}{\textit{e.g.,}\xspace}

\newcommand{\etal}{\textit{et al.}\xspace}

\newcommand{\secref}[1]{Section~\ref{#1}\xspace}

\newcommand{\figref}[1]{Figure~\ref{#1}\xspace}

\newcommand{\tabref}[1]{Table~\ref{#1}\xspace}

\newcommand{\revise}[1]{\textcolor{black}{#1}}
    
\begin{document}

\title{Investigating the Impact of Vocabulary Difficulty and Code Naturalness on Program Comprehension}

\author{\IEEEauthorblockN{Bin Lin}
\IEEEauthorblockA{\textit{Radboud University} \\
Nijmegen, the Netherlands\\
bin.lin@ru.nl}
\and
\IEEEauthorblockN{Gregorio Robles}
\IEEEauthorblockA{\textit{Universidad Rey Juan Carlos}\\
Madrid, Spain \\
grex@gsyc.urjc.es}
}

\maketitle

\begin{abstract}
Context: Developers spend most of their time comprehending source code during software development. Automatically assessing how readable and understandable source code is can provide various benefits in different tasks, such as task triaging and code reviews. While several studies have proposed approaches to predict software readability and understandability, most of them only focus on local characteristics of source code. Besides, the performance of understandability prediction is far from satisfactory.

Objective: In this study, we aim to assess readability and understandability from the perspective of language acquisition. More specifically, we would like to investigate whether code readability and understandability are correlated with the naturalness and vocabulary difficulty of source code. 

Method: To assess code naturalness, we adopted the cross-entropy metric, while we use a manually crafted list of code elements with their assigned advancement levels to assess the vocabulary difficulty. We will conduct a statistical analysis to understand their correlations and analyze whether code naturalness and vocabulary difficulty can be used to improve the performance of code readability and understandability prediction methods. The study will be conducted on existing datasets.
\end{abstract}



\section{Introduction}
\label{sec:intro}

Program comprehension is a vital activity during software development, as a study suggests that developers spend around 70\% of their time on it \cite{MinelliML15}. Easy-to-comprehend source code provides several benefits, such as facilitating code review and accelerating the on-boarding process for newcomers in the project. Over the past years, researchers have approached software comprehensibility from different perspectives, including readability and understandability. Readability refers to the legibility of a text, while understandability refers to how easy a text can be understood \cite{rello2013frequent}. Scalabrino \etal highlighted the difference between these two concepts in the software domain, with an example that a developer might find some well-structured code readable but still struggles to understand the code because of the unknown APIs used \cite{scalabrino_automatically_2021}.

Automatically assessing how readable and understandable a piece of code is can be leveraged in various scenarios. For example, developers may be asked to further improve their code when it is assessed as not readable/understandable before submitting the code for review. Another use case is that issues involving easy-to-understand source code can be assigned to junior developers in the project. Code with high readability and understandability can be also provided as examples to students in programming courses. Given these benefits, many studies have been conducted to predict the readability \cite{buse_learning_2010,posnett_simpler_2011,dorn2012general,scalabrino_improving_2016} and understandability \cite{scalabrino_automatically_2021, trockman_automatically_2018}.

However, the approaches proposed in these studies have certain limitations. First of all, most features used for prediction are local (\eg code complexity metrics, existence of comments), without considering the relation with the other software projects. Secondly, in most cases, no correlation has been found between the metrics considered in previous studies and code understandability~\cite{scalabrino_automatically_2021}. 

In this study, we take a different perspective when looking into readability and understandability. Think about our natural language acquisition process. When we are reading an article with a large amount of advanced vocabulary we rarely use in our daily life or we have never seen before, it is difficult to understand what message the article wants to convey. In fact, studies in linguistics have already confirmed this phenomenon: the difficulty of the vocabulary used has a direct and consistent impact on reading comprehension performance, regardless of the reader's ability \cite{freebody1983effects}. The language input repetition also has great effect on comprehension \cite{jensen2003exact}.  

We conjecture that program comprehension shares similarities with (natural) language acquisition: when source code uses more advanced elements or is written in a way less frequently seen in other software repositories, it might be more difficult to comprehend. In this registered report, we propose to analyze the correlation between code readability/understandability and two characteristics of software: 1) vocabulary difficulty, namely the amount of advanced code elements used; and 2) input repetition (software naturalness), namely how repetitive code elements are in software repositories. We plan to further validate whether these two elements can help improve the performance of readability and understandability prediction.

To ensure our results are comparable with previous studies, our study will focus on Java projects. The analysis will be conducted on existing datasets. \revise{We envision our study can benefit the software engineering academic and practitioner communities as follows: 
\begin{itemize}
    \item Our study is the first to establish links between code readability / understandability and non-local features (\ie vocabulary difficulty and code naturalness).
    \item The results can provide new insights on automatically assessing code readability and understandability.
    \item We address readability and understandability separately, which highlights the differences between these two concepts and can provide valuable experience for researchers working on code comprehension. 
    \item We will use a set of different datasets for evaluation, which can offer a more comprehensive view on how readability and understandability prediction models work.
\end{itemize}
}


\section{Related Work}
\label{sec:related}


\subsection{Code Readability}

Code readability can be impacted by different factors. For instance, Johnson \etal \cite{lee2013study} conducted a controlled experiment to understand the impact of nesting and do-while loops on code readability. Their results indicate that minimizing nesting does have a positive effect on reducing the effort needed for code reading by developers, while avoiding the do-while statement does not have a significant impact. Dos Santos and Gerosa \cite{dos2018impacts} analyzed the correlation between a set of 11 Java coding practices and code readability perceived by developers. Their results disclose that not all the coding practices can help improve the readability. 

Buse and Weimer~\cite{buse_learning_2010} constructed the first general model for predicting the readability of source code. Their model mainly relies on local code features (\eg line length, number of identifiers, identifier length). The performance of the model was evaluated on a dataset consisting of 100 code snippets, for which readability was manually scored by 120 human annotators. Their results indicate that the model can correctly predict readability for around 80 percent of the code snippets. They also found that readability is highly correlated with some metrics of software quality including defect density. However, code readability is not closely related to the cyclomatic complexity of source code. 

Posnett \etal \cite{posnett_simpler_2011} built upon the work of Buse and Weimer~\cite{buse_learning_2010} and presented a new model based on size and code entropy. Given a certain size, a higher entropy leads to higher readability, and vice versa. While being simpler, the model performs better and is better founded theoretically. 

Dorn \cite{dorn2012general} argued that code readability is highly relevant to how humans read on screens and integrated multiple visual, spatial, and linguistic features into the readability prediction model, such as structural patterns and sizes of code blocks. A new dataset containing 360 code snippets was created involving over 5,000 human evaluators. The evaluation shows that visual and spatial features can significantly improve the performance of the readability prediction model.

Scalabrino \etal \cite{scalabrino_comprehensive_2018} improves the readability model with a set of new textual features, including comments and identifiers consistency, identifier terms in dictionaries, narrow meaning identifiers, textual coherence, comments readability, and number of meanings. The performance was assessed on over 600 manually evaluated snippets, and the results show that textual features can be used to improve the readability prediction model. After analyzing 20 open-source Java projects, they found that the predictions of the new model better correlates with FindBugs warnings compared to previous models.  

Fakhoury \etal \cite{fakhoury2019improving}, nevertheless, disclosed the limitations of existing code readability metrics. They analyzed 548 readability commits from 63 Java projects, in which developers explicitly mentioned that the change was made to improve readability. Their results indicate that existing readability models are not able to capture readability improvements. 

Given the wide adoption of deep learning techniques in software engineering applications, Mi \etal \cite{mi_improving_2018} have also proposed to use 
Convolutional Neural Networks (ConvNets) to predict code readability. Their approach treats readability prediction as a binary classification task (\ie classifying a piece of code as readable or not readable). The evaluation results demonstrate that their approach outperforms previous models by 2.4 - 17.2\%.

\subsection{Code Understandability}

B{\"o}rstler and Paech \cite{borstler2016role} analyzed the impact of method chains and code comments on software comprehension. Their study involved 104 students and the metrics considered included the perceived readability, reading time and performance on a simple cloze test. Their results indicate that method chain or comment variants do not play a significant role for code comprehension. Moreover, the perceived readability is not correlated with comprehension measured by answer accuracy, which also highlights the difference between readability and understandability. 

Peitek \etal \cite{peitek2018look} investigated whether functional magnetic resonance imaging (fMRI) can be used to measure program comprehension. The study involving 17 participants with an fMRI scanner demonstrates that distinct cognitive processes can be identified during comprehension tasks, and the  observed brain activities can be used to understand the developers' program comprehension process.

Scalabrino \etal \cite{scalabrino_automatically_2017} are among the first who aim to design metrics specifically for assessing the understandability of a given code snippet. They collected 324 evaluations from 46 participants to understand the perceived understandability. They then examined to which level  understandability is correlated to several types of metrics computed on code, documentation, and developers. Their results reveal that none of these metrics, including code readability and complexity, are able to accurately capture the code understandability. Clearly, better metrics and further studies are required to tackle this issue. 

Trockman \etal \cite{trockman_automatically_2018} revisited the work of Scalabrino et al.~\cite{scalabrino_automatically_2017}. Based on the assumption that combining multiple features would better capture code understandability, they reanalyzed the data from the study using different statistical techniques. The results show that some of the metric combinations do have correlations with understandability. As an extension of Scalabrino \etal \cite{scalabrino_automatically_2017}, the original authors \cite{scalabrino_automatically_2021} further confirmed the findings of Trockman \etal \cite{trockman_automatically_2018}. However, the performance of understandability predictions is still not good enough to be used in practice. They also expanded the dataset to further strengthen the validity of their findings. 

Campbell \cite{campbell2018cognitive} proposed a new metric named ``Cognitive Complexity'' to measure understandability. Like Cyclomatic Complexity, Cognitive Complexity considers decision points (\eg conditions, loops). Moreover, it also adds weights based on the nesting levels. 
Mu{\~n}oz Bar{\'o}n \etal \cite{munoz2020empirical} analyzed the correlation between Cognitive Complexity and understandability evaluations of 427 code snippets. Their results suggest that Cognitive Complexity is indeed correlated with comprehension time and subjective understandability ratings. However, the relation with correctness of comprehension tasks and physiological measures are mixed. 
Lavazza \etal \cite{lavazza2023empirical} also performed an empirical study to understand whether Cognitive Complexity can help better predict code understandability, and they found that Cognitive Complexity is similarly correlated to code understandability as traditional measures and that it does not bring significant improvement in terms of understandability prediction performance. 

\revise{Wagner and Wyrich \cite{wagner2021code} conducted a large-scale empirical study to understand the correlation between intelligence / personality and code comprehension performance. 135 students participated the study and were asked to answer
comprehension questions on code snippets. The results indicate that intelligence impacts code comprehension, while
personality traits do not have the same effect.}


\subsection{Gaps and Opportunities}
\label{sec:related_gaps}
\revise{Although research on readability and understandability exists, there are still gaps that can be seen as research opportunities. First, currently no strong correlation has been identified between the metrics used in prior studies and code understandability. Second, the existing understandability prediction models do not yield reliable results. Third, existing studies mostly focus on local features of code snippets such as local code complexity and comment quality, without taking into account external projects.}

\revise{To close these gaps, we plan to first analyze the correlation between non-local features (\ie vocabulary difficulty and code naturalness) and readability/understandability, and then build a new model integrating these features for readability and understandability prediction to investigate whether these features could help improve the performance of these models.}


\section{Research Questions and Hypotheses}
\label{sec:rqs}

The goal of this study is to investigate whether the vocabulary difficulty and software naturalness are correlated to code readability and understandability. Our intuition is that source code will be easier to read and understand when it uses easier vocabulary and is more natural (more repetitive/frequently seen in other software repositories). \revise{Investigating the correlation helps us better understand the code comprehension process of developers.} Therefore, we ask the following two questions: 

\begin{tcolorbox}
\textbf{RQ1:} \revise{Is vocabulary difficulty correlated to code readability and understandability?}
\end{tcolorbox}

We formalize our RQ1 in the following hypotheses:

\begin{description}[align=left] 
  \item [\textbf{H$_{1.1}$:}] Source code with a more difficult vocabulary has a lower readability. 
  \item [\textbf{H0$_{1.1}$:}] There is no significant difference in readability between code with vocabulary of different difficulty levels.
    \item [\textbf{H$_{1.2}$:}] Source code with a more difficult vocabulary has a lower understandability.
  \item [\textbf{H0$_{1.2}$:}] There is no significant difference in understandability between code with vocabulary of different difficulty levels.
\end{description}

\begin{tcolorbox}
  \textbf{RQ2:} \revise{Is software naturalness correlated to code readability and understandability?}
\end{tcolorbox}

We formalize our RQ2 in the following hypotheses:

\begin{description}[align=left] 
  \item [\textbf{H$_{2.1}$:}] Source code which is more natural has a higher readability.  
  \item [\textbf{H0$_{2.1}$:}] There is no significant difference in readability between code with different levels of naturalness.  
  \item [\textbf{H$_{2.2}$:}] Source code which is more natural has a higher understandability.  
  \item [\textbf{H0$_{2.2}$:}] There is no significant difference in understandability between code with different levels of naturalness.  
\end{description}

\revise{Automatically assessing code readability and understandability can be useful during code review and programming education. Toward the goal of automatically assessment}, we propose following two research questions: 

\begin{tcolorbox}
\textbf{RQ3:} \revise{Can vocabulary difficulty help improve the accuracy of models predicting code readability and understandability?}
\end{tcolorbox}
 
We formalize our RQ3 in the following hypotheses:

\begin{description}[align=left] 
  \item [\textbf{H$_{3.1}$:}] Integrating the vocabulary difficulty feature into the readability prediction model can improve performance. 
  \item [\textbf{H0$_{3.1}$:}] There is no significant difference in performance of readability prediction between models with and without the vocabulary difficulty feature.
  \item [\textbf{H$_{3.2}$:}] Integrating the vocabulary difficulty feature into the understandability prediction model can improve performance. 
  \item [\textbf{H0$_{3.2}$:}] There is no significant difference in performance of understandability prediction between models with and without the vocabulary difficulty feature. 
\end{description}

\begin{tcolorbox}
  \textbf{RQ4:} \revise{Can software naturalness help improve the accuracy of models predicting code readability and understandability?}
  \end{tcolorbox}

  We formalize our RQ4 in the following hypotheses:

  \begin{description}[align=left] 
    \item [\textbf{H$_{4.1}$:}] Integrating the software naturalness feature into the readability prediction model can improve  performance. 
    \item [\textbf{H0$_{4.1}$:}] There is no significant difference in performance of readability prediction between models with and without the software naturalness feature.
    \item [\textbf{H$_{4.2}$:}] Integrating the software naturalness feature into the understandability prediction model can improve performance. 
    \item [\textbf{H0$_{4.2}$:}] There is no significant difference in performance of understandability prediction between models with and without the software naturalness feature.  
  \end{description}

\section{Variables}
\label{sec:variables}

Our RQs are set up as a correlational study, \revise{which aims to assess the statistical relationship between different variables}.  

\subsection{Proxies for Vocabulary Difficulty and Code Naturalness}

\textbf{\textit{Vocabulary difficulty.}} This interval variable indicates the amount of advanced vocabulary used in a code snippet. We plan to measure it using our manually crafted vocabulary list containing code elements with assigned difficulty levels.  

\textbf{\textit{Code naturalness.}} This interval variable indicates how natural the source code is (how common the code elements appear in other software repositories). We plan to measure it using cross-entropy. 

Please refer to our execution plan (\secref{sec:execution}) for the detailed information on how to represent these variables.

\subsection{Proxies for Readability and Understandability}

The variables for RQ1 and RQ2 are \textbf{\textit{readability}} and \textbf{\textit{understandability}}. These two variables indicate how readable and how understandable source code is. They are presented by the human evaluations provided in the dataset.  

The variables for RQ3 and RQ4 are \textbf{\textit{performance of readability prediction}} and \textbf{\textit{performance of understandability prediction.}} They are measured by common performance metrics such as the MAE (Mean Absolute Error).

\begin{table*}[ht]
\centering
\caption{Datasets used in the study.}
\label{tab:dataset}
\begin{tabular}{lrrl}
\toprule
\textbf{study} & \textbf{\# code snippets} & \textbf{\# evaluations} & \textbf{data aspects} \\ \midrule
\multicolumn{4}{c}{datasets for readability}                                                           \\ \hline
Dorn \cite{dorn2012general}   &  120           &    25,932                    &                    perceived readability score (1-5)                 \\
\revise{B{\"o}rstler and Paech \cite{borstler2016role} }              &                  30         &            520             &  perceived readability score (1-5)                      \\
Scalabrino \etal \cite{scalabrino_improving_2016}               &             200              &   1,800                  &     perceived readability score (1-5)          \\ \hline
\multicolumn{4}{c}{datasets for understandability}                                                     \\ \hline
Scalabrino \etal \cite{scalabrino_automatically_2021}               &             50              &          444      &    \revise{time spent and correctness for solving comprehension tasks}    \\
B{\"o}rstler and Paech \cite{borstler2016role}               &                  30         &            520             &  \revise{time spent and correctness for solving comprehension tasks}                       \\
Peitek \etal \cite{peitek2018look}               &       23                    &       1,150                  &  \revise{time spent and correctness for solving comprehension tasks}                             \\ \bottomrule
\end{tabular}
\end{table*}

\section{Datasets}
\label{sec:datasets}

In this study, we will mainly focus on Java programs, as Java is one of the most popular languages in practice and the readability and understandability of Java snippets have been widely studied in literature. To address our RQs, we will use existing datasets provided by previous studies. A summary of the datasets can be found in \tabref{tab:dataset}.

\subsection{Datasets for Readability}
Three datasets will be adopted for readability. 
\begin{itemize}
    \item Dorn \cite{dorn2012general}: This dataset contains 120 CUDA, 120 Java, and 120 Python code snippets. Here, we only use the 120 Java snippets (obtained from 10 projects). \revise{During the dataset creation, the annotators were also asked to rate the readability of snippets on a 1–5 Likert scale from very unreadable (1) to very readable (5).}
    \item \revise{B{\"o}rstler and Paech \cite{borstler2016role}: This dataset contains 30 different snippets, more specifically, they are modified versions of 5 original snippets (6 variants per snippet). To annotate the readability of these snippets, participants were asked to read the snippets and rate the readability on a scale from 1 to 5.}
    \item Scalabrino \etal \cite{scalabrino_comprehensive_2018}: This dataset contains 200 sampled methods with a size between ten and 50 lines of code (including comments) from four open source Java projects. \revise{Similar to Dorn \cite{dorn2012general}, the readability of each method was rated from 1 to 5 by nine evaluators.}
\end{itemize}

All the datasets are provided in the replication package of original studies. While the link to Dorn's \cite{dorn2012general} dataset is broken, a copy can be found in the online appendix of Scalabrino \etal \cite{scalabrino_comprehensive_2018}, which will be used in our study.

\subsection{Datasets for Understandability}
Three datasets will be adopted for understandability. 
\begin{itemize}
    \item Scalabrino \etal \cite{scalabrino_automatically_2021}: This dataset contains 50 Java/Android methods with 50$\pm$20 ELOCs extracted from ten popular open-source projects. Participants had to choose whether they understood the method. If so, the method would be hidden and they had to answer three verification questions about the method. The time spent and the correctness were recorded. In total, 444 evaluations were collected.

    \item B{\"o}rstler and Paech \cite{borstler2016role}: \revise{Besides the perceived readability scores mentioned before, the authors also designed cloze tests, in which participants need to fill in blanked-out parts of code, to measure comprehension. The time needed to complete the survey and the accuracy were recorded.} In total, 104 participants provided 520 data points.

    \item Peitek \etal \cite{peitek2018look}: This dataset contains 23 code snippets. \revise{41 undergraduate computer science students were asked to determine the output of these snippets. The time and correctness were measured.}
\end{itemize}

All datasets are available in their online replication package. Code snippets from B{\"o}rstler and Paech \cite{borstler2016role} and Peitek \etal \cite{peitek2018look} are inside PDF files, which implies that they have to be extracted manually.

\section{Execution Plan}
\label{sec:execution}
\begin{figure*}[t]
    \centering
    \includegraphics[width=0.86\linewidth]{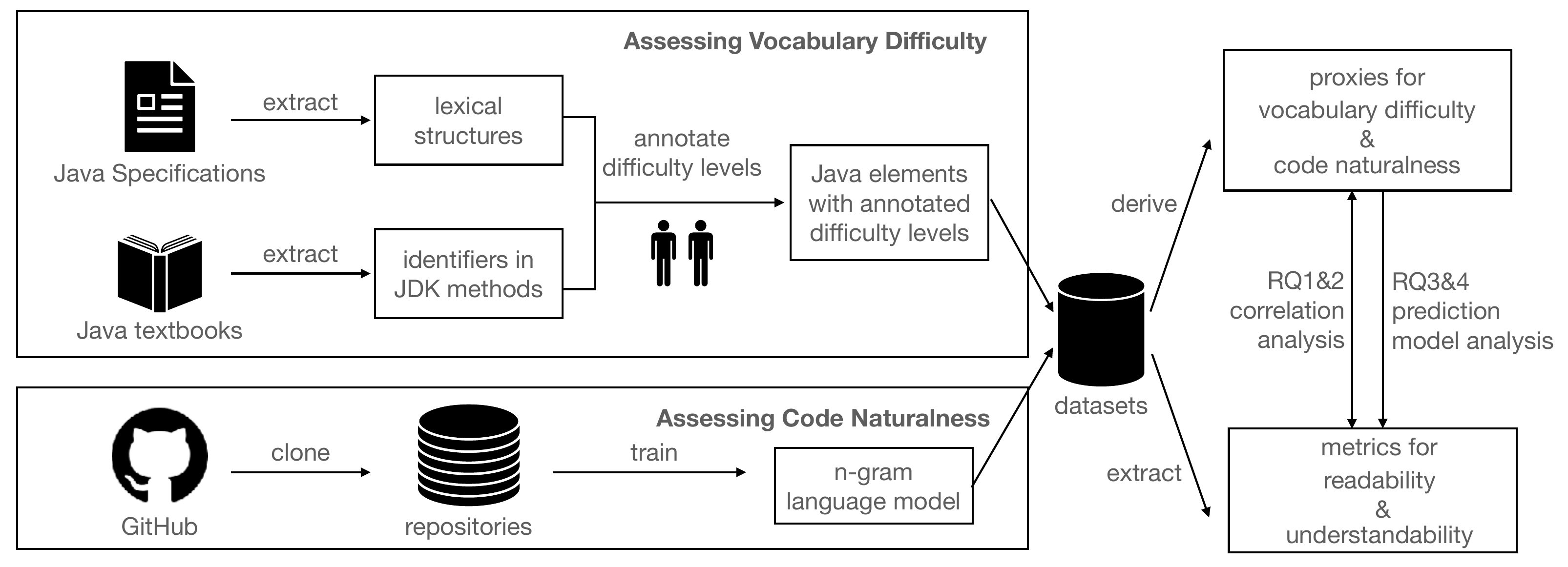}
    \caption{Overview of the execution plan}
    \label{fig:execution}
  \end{figure*}
  
In this section, we detail the plan on how we will execute the proposed study in this registered report. \revise{The overview of the  execution plan can be found in \figref{fig:execution}.}

\subsection{Assessing Vocabulary Difficulty}
\label{sec:voc_diff}

To assess vocabulary difficulty, we will adapt the approach applied in the study of Robles \etal \cite{robles2022a}. Following the well-established Common European Framework of Reference for Languages (CEFR) \cite{council2001common}, their study assigned a competency level (A1, A2, B1, B2, C1, or C2, where A1 is the lowest level and C2 is the highest level) to Python elements.

To create the list of Java elements with an assigned difficulty level, we will utilize popular Java textbooks (\ie \textit{Core Java, Volume I: Fundamentals, 12th Edition} \cite{corejava1} and \textit{Core Java, Volume II: Advanced Features, 12th Edition} \cite{corejava2}). More specifically, we will assign difficulty levels to Java elements from two sources:
\begin{enumerate}
    \item all the types of lexical structures defined in the Java language specification\footnote{\url{https://docs.oracle.com/javase/specs/jls/se17/html/jls-3.html}}, such as keywords (\eg \texttt{try}, \texttt{double}, \texttt{throw}), literals (\eg integer, Boolean, string), and operators (\eg \texttt{+}, \texttt{>=}, \texttt{>>}).
    
    \item all the identifiers in method calls related to JDK which appear in the code examples in these two textbooks. To extract these identifiers, we will use the symbol solver of JavaParser~\cite{smith2017javaparser} to extract all the fully-qualified names of method calls that appear in the code examples. For example, if \texttt{Math.min} is called in the code, we will retrieve \texttt{java.lang.Math.min(int, int)}. We will first check whether the called method is from JDK. We will ignore those which are not from JDK as when people start learning Java, their textbooks normally focus on how to use features provided by JDK instead of third-party libraries. Then we will tokenize the fully-qualified names and extract the identifiers (in this case \texttt{java}, \texttt{lang}, \texttt{Math}, \texttt{min}, \texttt{int}). 
\end{enumerate}

With these two lists of Java elements in hand, we will check in what chapter of the textbook the relevant elements are explained for the first time. Here we assume easier vocabulary will appear in earlier chapters, which corresponds to people's learning process. 
\revise{However, the difficulty level assignment does not need to strictly adhere to the order of vocabulary appearance in the textbook. For example, if the annotators think an element is easier than other elements of the same chapter, they can assign a lower difficulty level. We will create a web app for the annotation process, and the two authors will independently assign a difficulty level without knowing the assignment of the other annotator. Both annotators are university faculty members who have experience in teaching software development related courses, and are familiar with the Java programming language. To simplify the later calculation during our analysis, we will assign a number between 1 to 6 instead of A1-C2, where 1 represents A1 and 6 represents C2. The two authors will discuss the conflicts until an agreement is obtained. To reduce the bias during difficulty assignment and increase the level of agreement, we use adopt the following strategy: 
\begin{itemize}
    \item We will first randomly select 30 elements as our trial batch. The two authors will annotate these elements and discuss the disagreement before annotating the remaining elements. 
    \item We leverage the Java Certification Paths\footnote{\url{https://education.oracle.com/oracle-certification-path/pFamily_48}} of Oracle when determining the difficulty level. Oracle Corporation is the current owner of the Java SE platform and its Java certification program is the most widely accepted Java certification program. During the annotation, we will assign our elements covered by the syllabus of the ``Java Explorer'' course\footnote{\url{https://bit.ly/3DfBUuf}} to level A, assign elements which first appear in the ``Oracle Certified Associate'' program\footnote{\url{https://bit.ly/3rtkh7y}} to level B, and assign elements which first appear in the ``Oracle Certified Professional'' program\footnote{\url{https://bit.ly/3PS3KUU}} to level C. 
\end{itemize}
}   

We will use the following metrics as a proxy for the vocabulary difficulty level of a code snippet:
\begin{itemize}
    \item \# of elements in each difficulty level (NoLn, where n is an integer ranging from 1 to 6, \revise{\eg NoL1 of 20 means that 20 elements fall into difficulty level 1})
    \item ratio of elements in each difficulty level to the number of all the tokens (PerLn, where n is an integer ranging from 1 to 6, \revise{\eg PerL1 of 10\% means that 10\% of the elements fall into difficulty level 1}) 
    \item weighted sum of difficulty levels (the sum of PerLn * n, where n is an integer ranging from 1 to 6, \revise{\eg if PerL1 = 20\%, PerL2 = 80\%, and no element falls into other difficulty levels, weighted sum of difficulty levels would be 1*20\%+2*80\%=1.8)}  
    \item highest difficulty level in the code snippet (an integer ranging from 1 to 6, \revise{\eg a value of 5 means all the code elements have a difficulty level of 5 or below 5)}
\end{itemize}

\subsection{Assessing Code Naturalness}
\label{sec:ass_natural}

Since in linguistics it is clear that input repetition helps language comprehension \cite{jensen2003exact}, we assume that if a part of source code often appears in other software projects, it should be easier to comprehend as developers are more exposed to it. Indeed, software has been proven natural, namely very repetitive and predictable \cite{hindle2016naturalness}. The naturalness of software can be measured by cross-entropy. 

The \textit{cross-entropy} of a given code snippet $s$ composed by tokens $t_1...t_n$ of length $n$ is calculated as
$$ H_M\left(s\right) = -\frac { 1 }{ n } \log { P_M(s) } = -\frac { 1 }{ n }\sum _{ 1 }^{ n }{ \log { P_M\left( t_{ i }|h \right)  }  }  $$
where $P_M\left(s\right)$ is the probability of s estimated by the language model $M$, and $P\left( t_{ i }|h \right)$ is the probability of the token $t_i$ following its preceding tokens $h$ estimated by the language model $M$.

In our study, we will adopt the classic n-gram language model, which \revise{is used in the original paper measuring software naturalness~\cite{hindle2016naturalness}. The model} estimates the probability of having a token $t_i$ given the previous n-1 tokens, and assumes that the possibility of $t_i$ is only impacted by its preceding n-1 tokens: 
$p(t_i|t_{i-n+1}, ..., t_{i-2}, t_{i-1})$ =
$\frac{count(t_{i-n+1}t_{i-2} ... t_{i-1}t_i)}{count(t_{i-n+1}t_{i-2} ... t_{i-1}*)}$,
where * refers to any token.

It is worth noting that we will omit white spaces and comments when constructing the n-gram language models. Moreover, we will explore different variants of n-gram language models:
\begin{itemize}
    \item We will experiment with different lengths of n-gram (\ie 3, 5, 10). 
    \item We will experiment with n-grams with and without abstraction. The abstraction here refers to converting all the tokens to their lexical classes (\eg all the integers such as 1 and 100 will be converted to \texttt{INT}). 
\end{itemize}

To construct the n-gram language model, we will use all the \revise{Java projects from the Apache Software Foundation\footnote{\url{https://github.com/apache}}, the Eclipse Foundation\footnote{\url{https://github.com/eclipse}}, Android\footnote{\url{https://github.com/android}}, and Oracle\footnote{\url{https://github.com/oracle}} except those included in the datasets. We choose these foundations due to their profound impact in the Java programming community. We have not selected Java projects based on popularity metrics, such as stars and forks, since popular projects are often non-software-engineering projects, such as tutorials \cite{borges2018s}, and there are potential duplications that might introduce bias.} Although we do not use all the available Java projects on GitHub, given the large amount of Apache projects available (over 1000), we believe it will be enough to estimate how repetitive an unseen code snippet is compared to existing projects.  

\subsection{Obtain Readability}

As all the datasets were created in a similar fashion, we will merge the three datasets. The readability will also be represented with the averaged user ratings for the code snippet, which will range between 1 to 5. 

\subsection{Obtain Understandability}

As the datasets were created in different approaches and are also presented in different formats, we will discuss the results for each dataset individually. We will use the following metrics to represent understandability:
\begin{itemize}
    \item Time needed for comprehension: recorded time spent for addressing corresponding comprehension tasks
    \item Rate of correctness: percentage of correct solutions to the given tasks
\end{itemize}

\subsection{Analysis}
\label{sec:analysis} 

To address \revise{RQ1 and RQ2}, we will follow the approach adopted in the study of Scalabrino \etal \cite{scalabrino_automatically_2021}. That is, we will analyze which vocabulary difficulty/code naturalness metrics are correlated with which readability/understandability metrics. More specifically, we will use Kendall's rank correlation coefficient (\ie Kendall’s $\tau$) \cite{kendall1938new} as it does not assume that the data is normally distributed or a straight linear relationship exists between the analyzed pairs of metrics. To control the impact of multiple pairwise comparisons, we will adjust p-values with the Holm–Bonferroni method \cite{holm1979simple}. \revise{We will investigate the correlations of all the possible combinations of variable proxies, 4, 8, 1, and 2 pairs of proxies will be analyzed for H$_{1.1}$, H$_{1.2}$, H$_{2.1}$, and H$_{2.2}$, respectively. While several different proxies of vocabulary difficulty are involved, we do not expect all of them to be correlated with readability and understandability metrics. Our goal is to identify those proxies which are correlated or can help improve the performance of automatic assessment of readability and understandability.}

To address \revise{RQ3 and RQ4}, we will build a model for predicting metrics for readability and understandability. More specifically, we will examine whether integrating the features of vocabulary difficulty (\revise{measured by metrics defined in \secref{sec:voc_diff}}) and software naturalness (\revise{measured by code naturalness described in \secref{sec:ass_natural}})  can improve the performance of previous models. \revise{That is, we will compare the performance of existing models with and without our newly extracted features. The existing models have included various types of features, including code-related metrics (\eg lines of code, \# nested blocks, cyclomatic complexity) and documentation-related metrics (\eg comments and identifiers consistency, methods internal documentation quality).} As in Scalabrino \etal \cite{scalabrino_automatically_2021}, we will adopt several widely applied classifiers in the literature, including Linear Regression, Support Vector Machines, Neural Networks (Multilayer Perceptron), k-Nearest-Neighbors, and Random Forest. The correlation between the predicted and the actual values as well as the Mean Absolute Error (MAE) will be used as the evaluation metrics. Similarly we will also perform leave-one-out cross-validation, which means that for each fold we will have one instance for testing and the rest for training due to the relative small size of the datasets.

\subsection{Confounding Factor Analysis}
\label{sec:confounders}

\revise{Given the nature of our correlational study, no causation can be implied even if correlation is found. One reason is that some confounding factors might exist in our analysis. One common way to limit the effect of confounding factors is restriction, namely  eliminating variation in the confounder by selecting a comparison group \cite{pourhoseingholi2012control}. In this section, we will discuss how to use this approach to examine the impact of two important confounding factors. }

\revise{\textbf{\textit{Code length.}} The length of selected code snippets might impact the readability and understandability. We plan to eliminate the impact of code length by randomly select same number of continuous lines of code for all the code snippets, re-conduct the analysis and examine whether the correlational analysis results still hold.}

\revise{\textbf{\textit{\# comments.}} The existence of code comments might impact how developers interpret the code. We plan to conduct extra analyses to eliminate the impact of comments, that is, we will remove all comments of code snippets in the data and inspect whether the correlational relation remains unchanged. }


\section{Threats to Validity}
\label{sec:threats}

\emph{Threats to construct validity} concern the relationship between theory and observation. Program comprehension is a subjective process, and the evaluations provided by annotators can be biased due to their personal experiences. However, we believe using several datasets created by different groups of researchers can reduce this threat to a certain extent. 

\emph{Threats to internal validity} concern factors internal to our study that could have influenced our results. Given the variety of code snippets, it is impossible to completely rule out the possibility that the results are impacted by the confounding factors, unless we only use artificially modified snippets following a strict protocol. However, this will compromise how realistic our data is. In our study, we will perform extra analysis with controlled confounding variables to reduce such risks. We will also release our scripts and all the details to increase the transparency and replicability of our analysis.

\emph{Threats to external validity} concern the generalizability of our findings. Our study mainly focuses on readability and understandability of Java programs, therefore, it is unclear whether the results also apply to other programming languages. However, we believe most of the metrics we measure in this study also make sense in other languages, hence there is a high chance that the results can be generalized. Meanwhile, most of the snippets in the datasets are from open-source projects, which leads to the question whether the results remain unchanged for closed-source commercial software.

\section*{Replication}
\revise{To facilitate replication, we will release the our data and scripts on Zenodo\footnote{\url{https://zenodo.org/}}.}  

\balance

\bibliographystyle{IEEEtran} 
\bibliography{references}

\end{document}